# Feasible Automata for Two-Variable Logic with Successor on Data Words[*]

Ahmet Kara[1], Thomas Schwentick[1], and Tony Tan[2]

[1] Technical University of Dortmund
[2] University of Edinburgh

**Abstract.** We introduce an automata model for data words, that is words that carry at each position a symbol from a finite alphabet and a value from an unbounded data domain. The model is (semantically) a restriction of data automata, introduced by Bojanczyk, et. al. in 2006, therefore it is called *weak data automata*. It is strictly less expressive than data automata and the expressive power is incomparable with register automata. The expressive power of weak data automata corresponds exactly to existential monadic second order logic with successor $+1$ and data value equality $\sim$, $\mathsf{EMSO}^2(+1, \sim)$. It follows from previous work, David, et. al. in 2010, that the nonemptiness problem for weak data automata can be decided in 2-NEXPTIME. Furthermore, we study weak Büchi automata on data $\omega$-strings. They can be characterized by the extension of $\mathsf{EMSO}^2(+1, \sim)$ with existential quantifiers for infinite sets. Finally, the same complexity bound for its nonemptiness problem is established by a nondeterministic polynomial time reduction to the nonemptiness problem of weak data automata.

## 1 Introduction

Motivated by challenges in XML reasoning and infinite-state Model Checking, an extension of strings and finitely labelled trees by data values has been investigated in recent years. In classical automata theory, a string is a sequence of positions that carry a symbol from some finite alphabet. In a nutshell, *data strings* generalize strings, in that every position additionally carries a data value from some infinite domain. In the same way, *data trees* generalize (finitely) labelled trees. In XML Theory, data trees model XML documents. Here, the data values can be used to represent attribute values or text content. Both, cannot be adequately modelled by a finite alphabet. In a Model Checking[3] scenario, the data values can be used, e.g., to represent process id's or other data.

Early investigations in this area usually considered strings over an "infinite alphabet", that is, each position only have a value, but no finite-alphabet symbol [2,19,7,14,15,17]. Many of the automata models and logics that have been studied for data strings and trees lack the usual nice decidability properties of automata over finite alphabets, unless strong restrictions are imposed [10,4,3,1].

---

[*] The first and the second authors acknowledge the financial support by the German DFG under grant SCHW 678/4-1, and the third author Future and Emerging Technologies (FET) programme within the Seventh Framework Programme for Research of the European Commission, under the FET-Open grant agreement FOX, number FP7-ICT-233599.

[3] In the Model Checking setting, a position might carry a finite set of propositional variables, instead of a symbol.

A result that is particularly interesting for our investigations is the decidability of the satisfiability problem for two-variable logic over data strings [4]. Here, as usual, the logical quantifiers range over the positions of the data string and it can be checked whether a position $x$ carries a symbol $a$ (written: $a(x)$), whether it is to the left of a position $y$ ($x + 1 = y$), whether $x$ is somewhere to the left of $y$ ($x < y$) and whether $x$ and $y$ carry the same data value ($x \sim y$). The logic is denoted by $\mathsf{FO}^2(+1, <, \sim)$. The result was shown with the help of a newly introduced automata model for data words, *data automata* (DA). It turned out, that the expressive power of these automata can be actually characterized by the extension of $\mathsf{FO}^2(+1, <, \sim)$ with existential quantification over sets (of positions) and an additional predicate that holds for $x$ and $y$ if $y$ is the next position from $x$ with the same data value.

However, the complexity of the decision procedure for $\mathsf{FO}^2(+1, <, \sim)$ is very high. The problem is equivalent to the Reachability problem for Petri nets [12], a notoriously hard problem whose complexity has not been resolved exactly. Thus, it has been investigated how the complexity can be reduced, by dropping one of the predicates $x < y$ or $x + 1 = y$. In the latter case (that is, for $\mathsf{FO}^2(<, \sim)$) the complexity decreases to NEXPTIME [4]. In the former case ($\mathsf{FO}^2(+1, \sim)$) the complexity also becomes elementary. In [3] a 3-NEXPTIME bound was shown for the case of data trees and this bound clearly carries over to data strings. A more direct proof with a 4-NEXPTIME bound was given in [8] and a 2-NEXPTIME bound was obtained in [18].

The high complexity of the satisfiability of $\mathsf{FO}^2(+1, <, \sim)$ in [4] results from the high complexity of the nonemptiness problem for data automata. One of the starting questions for this paper was:

(1) Is there a natural restriction of data automata with (i) a better complexity and (ii) a correspondence to $\mathsf{FO}^2(+1, \sim)$?

We show that such a restriction indeed exists. Data automata consist of two automata $\mathcal{A}$ and $\mathcal{B}$. $\mathcal{A}$ is a non-deterministic letter-to-letter transducer that constructs, given the finite alphabet part of the input data string[4] $u$, a new data string $w$ (where, for each position, the data value in $w$ is the same as in $u$). The second automaton $\mathcal{B}$ can then check properties of the subsequences of $w$ that carry the same data value. We define *weak data automata (WDA)* which also use a non-deterministic letter-to-letter transducer but can only test some simple constraints of the subsequences in the second part. These constraints are (unary) key, inclusion and denial constraints and they are evaluated for each class separately (there are no inter-class constraints).

It turns out that WDA are expressively weaker than data automata, incomparable with register automata [14,1] and that their expressiveness can be precisely characterized by the extension of $\mathsf{FO}^2(+1, \sim)$ by existential set quantification, that is, $\mathsf{EMSO}^2(+1, \sim)$. As the property that we use to separate the expressive power of WDA and DA can be defined in $\mathsf{EMSO}^2(+1, <, \sim)$ we get that $\mathsf{EMSO}^2(+1, \sim) \not\equiv \mathsf{EMSO}^2(+1, <, \sim)$ as opposed to the classical setting (without data values) where $\mathsf{EMSO}^2(+1) \equiv \mathsf{EMSO}^2(+1, <)$. Indeed, one of the benefits of the logical characterization is that it gives an easy means to show non-expressibility for $\mathsf{EMSO}^2(+1, \sim)$ (and $\mathsf{FO}^2(+1, \sim)$). From results in [8] it immediately follows that the nonemptiness problem for WDA can be solved in 2-NEXPTIME.

---

[4] The transducer also sees whether a position has the same data value as the next one.

As mentioned above, one motivation to study data strings comes from Model Checking. In that context, systems are usually considered to run forever and to produce infinite traces. Thus, data $\omega$-words need to be considered as well, and this was actually one of the main motivations of this research. In particular we address the following questions.

(2) Do the complexity results of [8] carry over to data $\omega$-strings?
(3) Can the expressibility results and logical characterizations of the first part of the paper also be established for data $\omega$-strings?

It is straightforward to adapt weak data automata for data $\omega$-strings. The transducer can simply be equipped with a Büchi acceptance mechanism. We refer to the resulting model as *weak Büchi data automata (WBDA)*. It turns out that the answer to both questions, (2) and (3), is affirmative. For (3), this is not hard to prove. The separation of WDA from DA also separates WBDA from Büchi data automata. It is also not too hard to get a logical characterization of WBDA by extending $\mathsf{EMSO}^2(+1, \sim)$ with existential set quantifiers that are semantically restricted to bind to infinite sets. The answer to question (2) required considerably more effort. However, we establish a 2-NEXPTIME upper bound for the nonemptiness problem for WBDAs by a nondeterministic polynomial time reduction to the nonemptiness for WDA.

*Related work.* Some related work was already mentioned above. The pioneering works in Linear Temporal Logic for $\omega$-words with data are the papers [10,9]. In [9] an extension of Linear Temporal Logic (LTL) to handle data values is proposed and its satisfiability problem is shown to be decidable. The decision procedure is a reduction to the reachability problem in Petri nets, thus resulting in a similarly unknown complexity as for data automata. The logic and automata considered in [10] are decidable for finite data words, but not primitive recursive, and undecidable for $\omega$-words. In [16] it is shown that with a *safety* restriction both the logic and the automata become decidable, even in EXPSPACE. In [9] a logic with PSPACE complexity is considered. In [5], MSO logic on data words (with possibly multiple data values per position) is compared to automata models for various types of successor relations.

*Organization.* We give basic definitions in Section 2. In Section 3, weak data automata are defined, their complexity is given, and their expressive power is compared with other models. Section 4 gives the logical characterization of WDA by $\mathsf{EMSO}^2(+1, \sim)$. Section 5 studies data $\omega$-strings and shows how the nonemptiness problem of WBDA can be nondeterministically reduced in polynomial time to the nonemptiness of WDA. Section 6 states some open problems. .

*Acknowledgement.* We thank Christof Löding for helpful remarks on automata and logics for $\omega$-words and Thomas Zeume for thorough proof reading.

## 2 Notation

**Data words.** Let $\Sigma$ be a finite alphabet and $\mathfrak{D}$ an infinite set of data values. A *finite* word is an element of $\Sigma^*$, while an $\omega$-word is an element of $\Sigma^\omega$. A finite *data word* is an element of $(\Sigma \times \mathfrak{D})^*$, while a *data $\omega$-word* is an element of $(\Sigma \times \mathfrak{D})^\omega$. We often refer to data words also as *data strings*.

We write a data (finite or $\omega$-) word $w$ as $\binom{a_1}{d_1}\binom{a_2}{d_2}\cdots$, where $a_1, a_2, \ldots \in \Sigma$ and $d_1, d_2, \ldots \in \mathfrak{D}$. The symbol $a_i$ is the label of position $i$, while the value $d_i$ is the data value of position $i$. The projection of $w$ to the alphabet $\Sigma$ is denoted by $\mathsf{Str}(w) = a_1 a_2 \ldots$. A position in $w$ is called an *a-position*, if the label of that position is $a$. We denote by $V_w(a)$, the set of data values found in $a$-positions in $w$, i.e., $V_w(a) = \{d_i \mid a_i = a\}$, for each $a \in \Sigma$. Note that some $V_w(a)$'s may be infinite, while some others finite.

A maximal set of positions with the same data value $d$ is called a *class $c^d$* of the word and the $\Sigma$-string induced by the symbols at its positions is called the *class string $w^d$*. The *profile word* of a data $\omega$-word $w = \binom{a_1}{d_1}\binom{a_2}{d_2}\cdots$ is $\mathsf{Profile}(w) = (a_1, s_1), (a_2, s_2), \ldots \in (\Sigma \times \{\top, \bot\})^\omega$, where for each position $i \geq 1$ the component $s_i$ is $\top$ if and only if $d_i = d_{i+1}$. The profile word of a finite data word $\binom{a_1}{d_1}\binom{a_2}{d_2}\cdots\binom{a_n}{d_n}$ is defined similarly, with the addition that the component $s_n$ is $\bot$.

*Automata and Büchi automata.* An *automaton* $\mathcal{A}$ over the alphabet $\Sigma$ is a tuple $\mathcal{A} = \langle \Sigma, Q, q_0, \Delta, F \rangle$, where $Q$ is a finite set of states, $q_0 \in Q$ is the initial state, $\Delta \subseteq Q \times \Sigma \times Q$ is a set of transitions and $F \subseteq Q$ is a set of accepting states. A run of $\mathcal{A}$ on a word $w = a_1 a_2 \ldots a_n$ is a sequence $\rho = q_1 \ldots q_n$ of states from $Q - \{q_0\}$ such that $(q_0, a_1, q_1) \in \Delta$ and $(q_i, a_{i+1}, q_{i+1}) \in \Delta$ for each $1 \leq i < n$. The run $\rho$ is accepting, if $q_n \in F$.

A *Büchi automaton* $\mathcal{A}$ is syntactically just an automaton. A run of $\mathcal{A}$ on an $\omega$-word $w = a_1 a_2 \ldots$ is an infinite sequence $\rho = q_1 q_2 \ldots$ of states from $Q - \{q_0\}$ such that $(q_0, a_1, q_1) \in \Delta$ and $(q_i, a_{i+1}, q_{i+1}) \in \Delta$, for each $i \geq 1$. Let $\mathsf{Inf}(\rho)$ denote the set of states that appear infinitely many times in $\rho$. The run $\rho$ is accepting if $\mathsf{Inf}(\rho) \cap F \neq \emptyset$.

A word (resp. an $\omega$-word) $w$ is accepted by an automaton (resp. Büchi automaton) $\mathcal{A}$, if there exists an accepting run of $\mathcal{A}$ on $w$. As usual, $\mathcal{L}(\mathcal{A})$ (resp. $\mathcal{L}^\omega(\mathcal{A})$) denotes the set of words (resp. $\omega$-words) accepted by the automaton $\mathcal{A}$.

*Letter-to-letter transducers.* A *letter-to-letter transducer* over the input alphabet $\Sigma$ and the output alphabet $\Gamma$ is a tuple $\mathcal{T} = \langle \Sigma, \Gamma, Q, q_0, \Delta, F \rangle$, where $Q$, $q_0$, $F$ are the set of states, the initial state, and the set of accepting states, respectively, and $\Delta \subseteq Q \times \Sigma \times Q \times \Gamma$ is the set of transitions. The intuitive meaning of a transition $(q, a, q', \gamma)$ is that when the automaton is in state $q$, reading the symbol $a$, then it can move to the state $q'$ and output $\gamma$. A *run* of $\mathcal{T}$ on a word $w = a_1 a_2 \ldots a_n$ is a sequence $(q_1, \gamma_1), \ldots, (q_n, \gamma_n)$ over $(Q - \{q_0\}) \times \Gamma$ such that $(q_0, a_1, q_1, \gamma_1) \in \Delta$ and $(q_i, a_{i+1}, q_{i+1}, \gamma_{i+1}) \in \Delta$, for each $1 \leq i < n$. Likewise, a *run* of $\mathcal{T}$ on an $\omega$-word $w = a_1 a_2 \ldots$ is a sequence $(q_1, \gamma_1), (q_2, \gamma_2), \ldots$ over $(Q - \{q_0\}) \times \Gamma$ such that $(q_0, a_1, q_1, \gamma_1) \in \Delta$ and $(q_i, a_{i+1}, q_{i+1}, \gamma_{i+1}) \in \Delta$, for each $i \geq 1$. A run is *accepting* if it is accepting in the sense of (Büchi) automata. We say that $v = \gamma_1 \gamma_2 \ldots$ is an output of $\mathcal{T}$ on $w$, if there exists an accepting run $(q_1, \gamma_1), (q_2, \gamma_2), \ldots$ of $\mathcal{T}$ on $w$.

*Data automata.* A *data automaton (DA)* is a pair $(\mathcal{A}, \mathcal{B})$, where

- $\mathcal{A}$ is a letter-to-letter transducer with input alphabet $\Sigma \times \{\top, \bot\}$ and output alphabet $\Gamma$,
- $\mathcal{B}$ is a finite state automaton over the alphabet $\Gamma$.

A data word $w$ is accepted by $(\mathcal{A}, \mathcal{B})$ if the following holds.

– Profile($w$) is accepted by $\mathcal{A}$, yielding an output $u$.
  – For each data value $d$ of $w$, the class string $u^d$ is accepted by $\mathcal{B}$.

Data automata were introduced in the stated form in [4]. In [1] it was shown that their expressive power is not affected, if $\mathcal{A}$ gets $\mathsf{Str}(w)$ as input as opposed to $\mathsf{Profile}(w)$. In more recent papers, data automata are therefore defined in the (syntactically) weaker form with input $\mathsf{Str}(w)$.

## 3 Weak data automata

In this section we define a new automata model for finite data words and study its expressive power and its complexity. The model follows a similar approach as the model of data automata. The profile of the input data word is transformed by a letter-to-letter transducer and then further conditions on the resulting class strings are imposed. However, the conditions that can be stated in the new automata model are much more limited than those of a data automaton (hence the name *weak* data automata).

Let $\Gamma$ be an alphabet. Weak data automata allow three kinds of data constraints over $\Gamma$:

1. *key constraints*, written in the form: $\mathsf{key}(\gamma)$, where $\gamma \in \Gamma$.
2. *inclusion constraints*, written in the form: $V(\gamma) \subseteq \bigcup_{\gamma' \in R} V(\gamma')$, where $\gamma \in \Gamma$, $R \subseteq \Gamma$.
3. *denial constraints*, written in the form: $V(\gamma) \cap V(\gamma') = \emptyset$, where $\gamma, \gamma' \in \Gamma$.

Whether a data word $w$ satisfies a data constraint $C$, written as $w \models C$, is defined as follows.

1. $w \models \mathsf{key}(\gamma)$, if every two $\gamma$-positions in $w$ have different data values.
2. $w \models V(\gamma) \subseteq \bigcup_{\gamma' \in R} V(\gamma')$, if $V_w(\gamma) \subseteq \bigcup_{\gamma' \in R} V_w(\gamma')$.
3. $w \models V(\gamma) \cap V(\gamma') = \emptyset$, if $V_w(\gamma) \cap V_w(\gamma') = \emptyset$.

If $\mathcal{C}$ is a collection of data constraints, then we write $w \models \mathcal{C}$, if $w \models C$ for all $C \in \mathcal{C}$.

A *weak data automaton (WDA)* over the alphabet $\Sigma$ is a pair $(\mathcal{A}, \mathcal{C})$, where

  – $\mathcal{A}$ is a letter-to-letter transducer with input alphabet $\Sigma \times \{\top, \bot\}$ and output alphabet $\Gamma$,
  – $\mathcal{C}$ is a collection of data constraints over the alphabet $\Gamma$.

A data word $w = \binom{a_1}{d_1}\binom{a_2}{d_2} \cdots \binom{a_n}{d_n}$ is accepted by a WDA $(\mathcal{A}, \mathcal{C})$, if

  – there is an accepting run of $\mathcal{A}$ on $\mathsf{Profile}(w)$, with an output $\gamma_1 \gamma_2 \ldots \gamma_n$, and
  – the induced data word $w = \binom{\gamma_1}{d_1}\binom{\gamma_2}{d_2} \cdots \binom{\gamma_n}{d_n}$ satisfies all the constraints in $\mathcal{C}$.

We write $\mathcal{L}(\mathcal{A}, \mathcal{C})$ to denote the language that consists of all data words accepted by $(\mathcal{A}, \mathcal{C})$.

We first discuss some extensions of WDA by the constraints that were studied in [8].

  – *Disjunctive key constraints* are written in the form: $\mathsf{key}(K)$, where $K \subseteq \Gamma$. Such a constraint is satisfied by a data word if each of its classes has at most one position with a symbol from $K$.

- *Disjunctive inclusion constraints* are written in the form: $\bigcup_{\gamma \in S} V(\gamma) \subseteq \bigcup_{\gamma' \in R} V(\gamma')$, where $S, R \subseteq \Gamma$. Such a constraint is satisfied by a data word if each class with a position with a symbol from $S$ also has a position with a symbol from $R$.

An *extended weak data automaton* is defined like a WDA but it further allows disjunctive key and inclusion constraints.

**Lemma 1.** *From each extended WDA $(\mathcal{A}, \mathcal{C})$ an equivalent WDA of polynomial size can be constructed in polynomial time.*

*Proof.* A disjunctive inclusion constraint $\bigcup_{\gamma \in S} V(\gamma) \subseteq \bigcup_{\gamma' \in R} V(\gamma')$ can simply be replaced by a set of inclusion constraints $V(\gamma) \subseteq \bigcup_{\gamma' \in R} V(\gamma')$, one for each $\gamma \in S$. Disjunctive key constraints $\mathsf{key}(K)$ can be replaced by a set of denial constraints $V(\gamma) \cap V(\gamma') = \emptyset$, one for each pair $\gamma \neq \gamma'$ with $\gamma, \gamma' \in K$ and a set of key constraints $\mathsf{key}(\gamma)$, one for each $\gamma \in K$. □

Next, we compare the expressive power of weak data automata with other automata models for data words. More precisely we compare it with register automata [14,1] and data automata. Register automata are an extension of finite state automata with a fixed number of registers in which they can store data values and compare them with the data value of subsequent positions. For a precise definition we refer[1] the reader to [1].

We consider the following two data languages.

- $L_{a<b}$ consists of all data words over the alphabet $\{a, b\}$ with the property that for every $a$-position $i$ there is a $b$-position $j > i$ with the same data value;
- $L_{a*b}$ is the subset of $L_{a<b}$ where the next $b$-position $j$ with the same data value as $i$ always satisfies $j = i + 2$.

**Lemma 2.** *Neither $L_{a*b}$ nor $L_{a<b}$ can be decided by a WDA.*

*Proof.* We first show that no WDA decides $L_{a*b}$. Towards a contradiction, we thus assume that $L_{a*b}$ is decided by some weak data automata $(\mathcal{A}, \mathcal{C})$.

To this end, let $n = |\Gamma|^4 + 1$ and let $d_1, d'_1, d_2, d'_2, \ldots d_n, d'_n$ be pairwise different data values. We consider the data word

$$w = \binom{a}{d_1}\binom{a}{d'_1}\binom{b}{d_1}\binom{b}{d'_1}\binom{a}{d_2}\binom{a}{d'_2}\binom{b}{d_2}\binom{b}{d'_2}\cdots\binom{a}{d_n}\binom{a}{d'_n}\binom{b}{d_n}\binom{b}{d'_n}$$

of length $4n$. Clearly, $w$ is in $L_{a*b}$ and its profile is $((a, \bot)(a, \bot)(b, \bot)(b, \bot))^n$.

Let $\gamma = \gamma_1 \gamma_2 \cdots \gamma_{4n}$ be an output of $\mathcal{A}$ on the profile of $w$ such that $\binom{\gamma_1}{d_1} \cdots \binom{\gamma_{4n}}{d'_n}$ satisfies all constraints in $\mathcal{C}$. By the choice of $n$, there exist numbers $i, j$ with $0 \leq i < j < n$ such that $\gamma_{4i+1}\gamma_{4i+2}\gamma_{4i+3}\gamma_{4i+4} = \gamma_{4j+1}\gamma_{4j+2}\gamma_{4j+3}\gamma_{4j+4}$.

Let $u$ be the data word obtained from $w$ by swapping the positions of the data values $d_{i+1}d'_{i+1}$ and $d_{j+1}d'_{j+1}$. That is,

$$u = \binom{a}{d_1}\cdots\binom{a}{d_{i+1}}\binom{a}{d'_{i+1}}\binom{b}{d_{j+1}}\binom{b}{d'_{j+1}}\cdots\binom{a}{d_{j+1}}\binom{a}{d'_{j+1}}\binom{b}{d_{i+1}}\binom{b}{d'_{i+1}}\cdots\binom{b}{d'_n}.$$

---
[1] The precursor model *finite-memory automata* was defined on "strings over infinite alphabets", that is, essentially data strings without a $\Sigma$-component [14].

Clearly, $u \notin L_{a*b}$. However, because $\mathsf{Profile}(u) = \mathsf{Profile}(w)$, $\gamma_1 \gamma_2 \ldots \gamma_{4n}$ is also an output of $\mathcal{A}$ on $\mathsf{Profile}(u)$. Moreover, the sets of $V_u(\gamma) = V_w(\gamma)$, for each $\gamma \in \Gamma$, and therefore the validity of inclusion and denial constraints does not change. Furthermore, as in $u$ and $w$ every data value occurs at exactly one $a$-position and at exactly one $b$-position, they cannot be distinguished by key constraints, either. Thus, $u \in \mathcal{L}(\mathcal{A}, \mathcal{C})$, the desired contradiction.

The proof for $L_{a<b}$ is exactly the same, as $w \in L_{a<b}$ and $u \notin L_{a<b}$ (because of $\binom{a}{d_{j+1}}$). □

**Theorem 1.** *(a) The class of data languages that are decided by WDA is strictly included in the class of data languages decided by DA.*
*(b) The classes of languages decided by WDA and by register automata are incomparable.*

*Proof.* Towards (a) we first show that every WDA can be translated into a DA and thus WDA decide a subclass of DA. That the subclass is strict can then be concluded from (b) as register automata are captured by DA [1] and thus there is a data language that can be decided by a DA but not a WDA.

Let thus $(\mathcal{A}, \mathcal{C})$ be a WDA. Then $(\mathcal{A}, \mathcal{B})$ is a data automaton for $L(\mathcal{A}, \mathcal{C})$, where the automaton $\mathcal{B}$ tests the constraints in $\mathcal{C}$ as follows.

- For every key constraint $\mathsf{key}(\gamma)$ of $\mathcal{C}$, $\mathcal{B}$ tests that every class string has at most one $\gamma$-position.
- For every inclusion constraint $V(\gamma) \subseteq \bigcup_{\gamma' \in R} V(\gamma')$, $\mathcal{B}$ tests that every class string with a $\gamma$-position also has a $\gamma'$-position, for some $\gamma' \in R$.
- For every denial constraint $V(\gamma) \cap V(\gamma') = \emptyset$, $\mathcal{B}$ checks that classes with a $\gamma$-position do not have any $\gamma'$-positions.

To show statement (b) we first consider the separation language $L = L_{a*b}$ which cannot be decided by a WDA by Lemma 2. However, $L_{a*b}$ can be easily decided by a register automaton that always stores the last two data values in two registers and the information about their symbols in its state.

On the other hand, the set of all data strings over $\Sigma = \{a\}$ in which every data value occurs only once can easily be decided by a WDA by the identity-transducer and the key constraint $\mathsf{key}(a)$ but not by a register automaton [14]. □

The complexity of the nonemptiness problem for WDA follows directly from results in [8].

**Theorem 2.** *The nonemptiness problem for WDA is decidable in 2-NEXPTIME.*

*Proof.* In [8], it was shown that given an automaton $\mathcal{A}$ that reads profile strings and a set $\mathcal{C}$ of disjunctive key and inclusion constraints, to decide whether there is a data word $w$ such that $\mathcal{A}$ accepts $\mathsf{Profile}(w)$ and $w \models \mathcal{C}$ can be done in nondeterministic double exponential time.

Clearly, this is basically the same as the nonemptiness problem for WDA with disjunctive key and inclusion constraints only. It thus only remains to show that denial constraints can be translated into disjunctive constraints in a nonemptiness respecting fashion. To this end, a denial constraint $V(\gamma_1) \cap V(\gamma_2) = \emptyset$ can be replaced as follows. We add two new symbols $\gamma_1', \gamma_2'$ and require that in each class with $\gamma_i$ one $\gamma_i'$ occurs but $\gamma_1'$ and $\gamma_2'$ do not co-occur by two inclusion constraints $V(\gamma_1) \subseteq V(\gamma_1')$ and $V(\gamma_2) \subseteq V(\gamma_2')$ and a disjunctive key constraint for $\{\gamma_1', \gamma_2'\}$. □

## 4 A logical characterization of weak data automata

In this section, we give a logical characterization of the data languages decided by weak data automata in terms of existential second order logic. The characterization is an analogue of the Theorem of Büchi, Elgot and Trakhtenbrot [6,11,21] for string languages. This theorem can be stated for various logics, the most interesting one for our context is that $\mathsf{EMSO}^2(+1)$ characterizes exactly the regular languages.

We represent data words by logical structures $w = \langle \{1,\ldots,n\}, +1, <, \{a(\cdot)\}_{a\in\Sigma}, \sim \rangle$, where $\{1,\ldots,n\}$ is the set of positions, $+1$ is the successor relation (i.e., $+1(i,j)$ if $i+1=j$), $<$ is the order relation (i.e., $<(i,j)$ if $i<j$), the $a(\cdot)$'s are the label relations, and $i \sim j$ holds if positions $i$ and $j$ have the same data value. As the empty data word can not be properly represented, the logical characterization of WDA ignores empty data words. That is, if some WDA $(\mathcal{A}, \mathcal{C})$ accepts the empty data string then its language is different from the language of the corresponding formula $\varphi$: $\mathcal{L}(\mathcal{A}, \mathcal{C}) = L(\varphi) \cup \{\epsilon\}$.

For a set $SS \subseteq \{+1, <, \sim\}$ of relation symbols, we write $\mathsf{FO}(SS)$ for first-order logic with the vocabulary $SS$, $\mathsf{MSO}(SS)$ for monadic second-order logic (which extends $\mathsf{FO}(SS)$ with quantification over sets of positions), and $\mathsf{EMSO}(SS)$ for existential monadic second order logic, that is, all sentences of the form $\exists R_1 \ldots \exists R_m\, \psi$, where $\psi$ is an $\mathsf{FO}(SS)$ formula extended with the unary predicates $R_1, \ldots, R_m$. By $\mathsf{FO}^2(SS)$ we denote the restriction of $\mathsf{FO}(SS)$ to sentences with two variables $x$ and $y$, and by $\mathsf{EMSO}^2(SS)$ the restriction of $\mathsf{EMSO}(SS)$ where the first-order part uses only two variables.

### 4.1 From weak data automata to $\mathsf{EMSO}^2(+1, \sim)$

**Theorem 3.** *For every weak data automaton $(\mathcal{A}, \mathcal{C})$, an equivalent $\mathsf{EMSO}^2(+1, \sim)$-formula $\varphi$ is constructible in polynomial time.*

*Proof.* Let $(\mathcal{A}, \mathcal{C})$ be a weak data automaton with $\mathcal{A} = \langle \Sigma, \Gamma, Q, q_0, \Delta, F \rangle$, where $Q = \{q_1, \ldots, q_n\}$ and $\Gamma = \{\gamma_1, \ldots, \gamma_l\}$. We recall that we assume without loss of generality that $\mathcal{A}$ uses $q_0$ only its initial state.

We will construct an $\mathsf{EMSO}^2(+1, \sim)$-formula $\varphi$ with $L(\mathcal{A}, \mathcal{C}) - \{\epsilon\} = L(\varphi)$. The construction is the same as the classical translation from NFAs to MSO formulas. See, for example, [20].

The formula $\varphi$ is

$$\varphi = \exists R_{q_1} \ldots \exists R_{q_n} \exists R_{\gamma_1} \ldots \exists R_{\gamma_l} (\varphi_{part} \wedge \varphi_{start} \wedge \varphi_{trans} \wedge \varphi_{accept} \wedge \varphi_{constr})$$

where

- $\varphi_{part}$ asserts in a straightforward manner that $R_{q_1}, \ldots, R_{q_n}$ on one hand and $R_{\gamma_1}, \ldots, R_{\gamma_l}$, on the other hand, partition the positions of the input word.
- $\varphi_{start}$ asserts that the automaton starts in state $q_0$:

$$\forall x (\neg \exists y\ x = y+1 \rightarrow \phi)$$

where $\phi$ is:

$$\forall y \ y = x+1 \to \bigwedge_{a \in \Sigma} \begin{pmatrix} (a(x) \land x \sim y) \to \bigvee_{(q_0,(a,\top),q,\gamma) \in \Delta} (R_q(x) \land R_\gamma(x))) \\ \land \\ (a(x) \land x \not\sim y) \to \bigvee_{(q_0,(a,\bot),q,\gamma) \in \Delta} (R_q(x) \land R_\gamma(x))) \end{pmatrix}$$

- $\varphi_{trans}$ asserts that transitions are simulated correctly:

$\forall x \forall y \ y = x+1 \to$

$$\bigwedge_{a \in \Sigma, q \in Q} \begin{pmatrix} (a(y) \land R_q(x) \land \exists x(x = y+1 \land y \sim x)) \to \bigvee_{(q,(a,\top),q',\gamma) \in \Delta} R_{q'}(y) \land R_\gamma(y) \\ \land \\ (a(y) \land R_q(x) \land \exists x(x = y+1 \land y \not\sim x)) \to \bigvee_{(q,(a,\bot),q',\gamma) \in \Delta} R_{q'}(y) \land R_\gamma(y) \end{pmatrix}$$

- $\varphi_{accept}$ states the accepting condition:

$$\forall x \big[ \neg \exists y \ y = x+1 \to \bigvee_{q \in F} R_q(x) \big]$$

- $\varphi_{constr}$ is a conjunction $\bigwedge_{C \in \mathcal{C}} \psi_C$ over all constraints $C$ from $\mathcal{C}$ such that,

  - if $C$ is a key constraint $\mathsf{key}(\gamma)$, then

  $$\psi_C = \forall x \forall y [(R_\gamma(x) \land R_\gamma(y) \land x \sim y) \to x = y]$$

  - if $C$ is an inclusion constraint $V(\gamma) \subseteq \bigcup_{\gamma' \in S} V(\gamma')$, then

  $$\psi_C = \forall x \exists y [R_\gamma(x) \to (\bigvee_{\gamma' \in S} R_{\gamma'}(y) \land x \sim y)]$$

  - if $C$ is a denial constraint $V(\gamma) \cap V(\gamma') = \emptyset$, then

  $$\psi_C = \forall x \forall y [(R_\gamma(x) \land R_{\gamma'}(y)) \to x \not\sim y]$$

The length of $\varphi$ is $\mathcal{O}(|\Sigma||Q||\Delta|+|\mathcal{C}|)$. The correctness is straightforward and, thus, omitted. □

## 4.2 From $\mathsf{EMSO}^2(+1, \sim)$ to weak data automata

In the following, we use the abbreviation $F(x,y)$ for the formula $\neg y = x+1 \land \neg x = y+1 \land x \neq y$, which states that the distance of $x$ and $y$ is at least two.

**Theorem 4.** *There is an algorithm that translates every $\mathsf{EMSO}^2(+1, \sim)$-formula $\varphi$ into an equivalent weak data automaton $(\mathcal{A}, \mathcal{C})$ in doubly exponential time. In particular, the output alphabet $\Gamma$ of $\mathcal{A}$ and the number of constraints in $\mathcal{C}$ is at most exponenotial.*

*Proof.* In the first step, the algorithm transforms $\varphi$ into an equivalent $\mathsf{EMSO}^2(+1, \sim)$ formula in Scott normal form (SNF) of the form

$$\psi = \exists R_1 \ldots \exists R_n \ [\forall x \forall y \ \chi' \wedge \bigwedge_{i=1}^{m} \forall x \exists y \ \chi'_i],$$

where $\chi'$ and each $\chi'_i$ are quantifier-free [13]. The size of $\psi$ is linear in the size of $\varphi$, in particular, $n = \mathcal{O}(|\varphi|)$ and $m = \mathcal{O}(|\varphi|)$.

Then it rewrites formula $\chi'$ into an, at most exponential, conjunction

$$\chi = \bigwedge_j \neg (\alpha_j(x) \wedge \beta_j(y) \wedge \delta_j(x, y) \wedge \epsilon_j(x, y)),$$

where, for every $j$, $\alpha_j, \beta_j$ are conjunctions of literals with unary relation symbols, $\delta_j$ is either $x \sim y$ or $x \not\sim y$ and $\epsilon_j(x, y)$ is one[2] of $x = y$, $y = x + 1$, $F(x, y)$.

Likewise, it rewrites every $\chi'_i$ into an, at most exponential, disjunction

$$\chi_i = \bigvee_j (\alpha^i_j(x) \wedge \beta^i_j(y) \wedge \delta^i_j(x, y) \wedge \epsilon^i_j(x, y)),$$

where the atomic formulas are of the respective forms as above.

The idea of the construction is that $\mathcal{A}$ guesses[3] a couple of relations that allow to state some of the properties expressed in $\psi$ by constraints of $\mathcal{C}$.

For simplicity we refer to the label relations $a_1, \ldots, a_l$ as $R_{n+1}, \ldots, R_{n+l}$.

The relations that are guessed are the following.

- $R_1, \ldots, R_n$ (the *SNF relations*). We refer to the full atomic type of a position with respect to the relations $R_1, \ldots, R_{n+l}$ as its *SNF-type*;
- $P_1, P_2, P_3$ with the following intention: if a class contains at least three positions of some SNF-type $\alpha$, then one of them is in $P_3$. If a class contains at least two $\alpha$-positions, then one of them is in $P_2$. If there is at least one $\alpha$-position then there is one in $P_1$.
- $C_1, C_2, C_3$ with the following intention: if there are at least three classes that contain positions of some SNF-type $\alpha$, then in one of these classes all $\alpha$-positions are in $C_3$. If there are at least two classes that contain $\alpha$-positions, then in one of them all $\alpha$-positions are in $C_2$. If there is at least one class with $\alpha$-positions then in one of them all $\alpha$-positions are in $C_1$. We refer to the full atomic type of a position with respect to $P_1, P_2, P_3, C_1, C_2, C_3$ as its *occurrence type*;
- $E_\leftarrow, E_\rightarrow$ with the intention that a position is in $E_\leftarrow$ if its left neighbor has the same data value (and likewise for $E_\rightarrow$);
- $R^\ell_1, \ldots, R^\ell_{n+l}, P^\ell_1, P^\ell_2, P^\ell_3, C^\ell_1, C^\ell_2, C^\ell_3$ and $R^r_1, \ldots, R^r_{n+l}, P^r_1, P^r_2, P^r_3, C^r_1, C^r_2, C^r_3$ with the following intention: For each position $p$, it should hold that $p$ is in a relation with superscript $\ell$, if its left neighbor is in the corresponding relation without superscript. Likewise, $p$ is in a relation with superscript $r$, if its right neighbor is in the corresponding relation without superscript. We refer to the type of a position with respect to these relation and the relations $E_\leftarrow, E_\rightarrow$ as its *neighborhood type*;

---

[2] The case $x = y + 1$ does not need to be considered as it can be obtained by swapping $x$ and $y$.
[3] More precisely, $\mathcal{A}$ guesses, for each position $p$, the set of those relations that contain $p$. However, on a global level, we refer to this as "guessing the relations".

- for every $j \leq n+l$ and $i \leq m$ the relations $W_j^i$ and $E^i, G_{suc}^i, G_{pre}^i, G_F^i$ with the following intention: if for some position $p$, formula $\chi_i$ becomes true for $x = p$ and some position $y = q$ then $p \in W_j^i$ if and only if $q \in R_j$. That is, $W_1^i, \ldots, W_{n+l}^i$ mimics the SNF-type of witness positions with respect to $\chi_i$. Furthermore, $p \sim q$ if and only if $p \in E^i$, $q = p+1$ if and only if $p \in G_{suc}^i$, $p = q+1$ if and only if $p \in G_{pre}^i$, and $F(p,q)$ if and only if $p \in G_F^i$. We refer to the type with respect to these relations as the *witness type*.

It should be noted that the number of these relations is $\mathcal{O}(nm)$.

Now we describe how $\mathcal{A}$ and $\mathcal{C}$ can be constructed in order to test whether a data string $w$ satisfies $\varphi$.

First, the automaton $\mathcal{A}$ guesses which types will occur in the output and verifies during its computation that exactly these types occur. What needs to verified is how $\mathcal{A}$ and $\mathcal{C}$ can ensure that the relations guessed by $\mathcal{A}$ are consistent with respect to the intention that was described above.

- The consistency with respect to $P_1$, $P_2$ and $P_3$ can be tested as follows. $\mathcal{A}$ can ensure that each position is in at most one $P_k$. The intention of $P_1$, $P_2$ and $P_3$ can be enforced by the the inclusion constraints
  - $V(\alpha \wedge \neg P_1 \wedge \neg P_2) \subseteq V(\alpha \wedge P_3)$,
  - $V(\alpha \wedge P_3) \subseteq V(\alpha \wedge P_2)$, and
  - $V(\alpha \wedge P_2) \subseteq V(\alpha \wedge P_1)$.

  That each class contains at most one position with $\alpha \wedge P_k$, for each $k$, can be stated by key constraints.
- The consistency with respect to $C_1$, $C_2$ and $C_3$ can be tested in a similar fashion. That, for some $\alpha$ the existence of an $(\alpha \wedge C_3)$-class implies the existence of an $(\alpha \wedge C_2)$-class and the other two corresponding conditions can be already guaranteed when $\mathcal{A}$ guesses the set of occurring types. That all $\alpha$-positions in an $(\alpha \wedge C_k)$-class are in $C_k$ can be ensured by denial constraints.
- The consistency of the neighborhood types can be easily tested by $\mathcal{A}$ with the help of the profile information.
- For the consistency of the witness types $\mathcal{A}$ checks that for each position $p$ and each $i \leq m$ there actually exists a disjunct $(\alpha_j^i(x) \wedge \beta_j^i(y) \wedge \delta_j^i(x,y) \wedge \epsilon_j^i(x,y))$ of $\chi_i$, for which $\alpha_j^i$ is the SNF-type of $p$ and $\beta_j^i$, $\delta_j^i$ and $\epsilon_j^i$ coincide with the $i$-th witness type of $p$. How the existence of corresponding witness positions is tested will be described below.

Now we describe how the conjuncts of $\chi$ can be checked. For every conjunct (indexed by $j$) we distinguish the following cases depending on the possible formulas $\delta_j(x,y)$ and $\epsilon(x,y)$.

(Case 1) $\epsilon$ is $x = y$: in this case the conjunct states a condition about forbidden SNF-types of positions. This kind of constraints can be ensured by $\mathcal{A}$ by not allowing to guess them.

(Case 2) $\epsilon$ is $y = x+1$: such conjuncts state that some pairs of SNF-types are forbidden for neighbors with equal (or different) data values. As this is a question of consistency between the neighborhood type and the witness type of a position it can be guaranteed by $\mathcal{A}$ (by disallowing certain combinations).

(Case 3) $\epsilon$ is $F(x,y)$ and $\delta$ is $x \sim y$: such a formula states that there should not be an $\alpha$-position $p$ and a $\beta$-position $q$ in the same class with $|p-q| > 1$. Such a formula gives rise to some denial constraints. As an example, if the SNF-type of a position $p$ is $\alpha$ and the neighborhood type of $p$ indicates that it has a $(P_1 \wedge \beta)$-position and a $(P_2 \wedge \beta)$-position as neighbors then there is a denial constraint forbidding $(P_3 \wedge \beta)$-positions in this class.

(Case 4) $\epsilon$ is $F(x,y)$ and $\delta$ is $x \nsim y$: such a formula states that $\alpha$-positions $p$ and $\beta$-positions $q$ with $|p-q| > 1$ need to be in the same class. This can be tested by $\mathcal{A}$ and some denial constraints with the help of $C_1, C_2, C_3$ and $P_1, P_2, P_3$.

What remains to be shown is how the formulas $\chi_i$ can be tested. For each position $p$ and each $i \le m$, $(\mathcal{A}, \mathcal{C})$ has to check that there is a witness position $q$ such that $p$ and $q$ satisfy some disjunct of $\chi_i$. Which disjunct should be considered is given by the $i$-th witness type of $p$. The existence of corresponding positions can be tested in a way that is similar to the tests for formula $\chi$. If $\epsilon$ is $x = y$ the witness needs to be $p$ itself which can be tested by $\mathcal{A}$ directly. If $\epsilon$ is $y = x+1$ or $x = y+1$ the witness is one of the neighbors of $p$. Whether this is true can be concluded from the neighborhood type and thus also by $\mathcal{A}$. If $\epsilon$ is $F(x,y)$ the existence of witnesses can be checked by inclusion constraints if $\delta$ is $x \sim y$ and by $\mathcal{A}$ directly if $\delta$ is $x \nsim y$. In the former case, the $P_k$-relations are used, in the latter case, both the $P_k$ and the $C_k$-positions.

The size of the output alphabet of $\mathcal{A}$ and the number of constraints are at most exponential in $|\varphi|$. The number of states is at most doubly exponential. □

We note that in the upper bound of the algorithm for nonemptiness of WDA transferred from [8], the doubly exponential term only depends on the alphabet size. By combing this with the bounds of Theorem 4 we obtain a 3-NEXPTIME upper bound for satisfiability of $\mathsf{FO}^2(+1, \sim)$ (which is worse than the bound in [18]). We also note that the construction underlying the proof of Theorem 4 can be turned into a nondeterministic exponential time reduction from satisfiability for $\mathsf{FO}^2(+1, \sim)$ to nonemptiness for WDA resulting in an automaton with a singly exponential number of states. The reduction guesses the order in which types appear in the accepted string (as opposed to the construction in the proof of Theorem 4).

The previous two theorems yield the following logical characterization of WDA.

**Theorem 5.** *Weak data automata and $\mathsf{EMSO}^2(+1, \sim)$ are equivalent in expressive power.*

We note that on strings $\mathsf{EMSO}^2(+1)$ and $\mathsf{EMSO}^2(+1, <)$ are expressively equivalent. It is an interesting consequence of the above characterization that this equivalence does not hold for data strings.

**Corollary 1.** *The logic $\mathsf{EMSO}^2(+1, \sim)$ is strictly less expressible than $\mathsf{EMSO}^2(+1, <, \sim)$.*

*Proof.* The inclusion holds by definition. It is strict because

- the language $L_{a<b}$ cannot be decided by a WDA (Lemma 2) and thus cannot be defined in $\mathsf{EMSO}^2(+1, \sim)$,
- but it can be expressed by the simple formula $\forall x \exists y (a(x) \to (b(y) \wedge x < y \wedge x \sim y))$.

□

## 5 Weak Büchi data automata

In this section we consider automata and logics for *data ω-words*, that is, data words of infinite length. Weak data automata $(\mathcal{A}, \mathcal{C})$ can easily be adapted for data ω-words. The automaton $\mathcal{A}$ is simply interpreted as a letter-to-letter Büchi transducer. A run is accepting if it visits infinitely often a state from $F$. We refer to the resulting model as weak Büchi data automata (WBDA). We write $\mathcal{L}^\omega(\mathcal{A}, \mathcal{C})$ for the set of data ω-words accepted by $(\mathcal{A}, \mathcal{C})$. The results regarding expressive power of WDA compared with other automata models easily carry over to WBDA.

Data ω-words can be represented by logical structures

$$w = \langle \mathbb{N}, +1, <, \{a(\cdot)\}_{a \in \Sigma}, \sim \rangle, \tag{1}$$

where $\mathbb{N}$ is the set $\{1, 2, \ldots\}$ of natural numbers which represent the positions and the other relations are as in the case of data words. For a set $SS \subseteq \{+1, <, \sim\}$ of relation symbols $\mathsf{E}_\infty \mathsf{MSO}(SS)$ consists of all formulas of the form

$$\exists_\infty R_1 \ldots \exists_\infty R_m \exists S_1 \ldots \exists S_\ell \ \varphi, \tag{2}$$

where $\varphi \in \mathsf{FO}^2(SS)$. Here all relation symbols $R_i, S_i$ are unary. The $\exists_\infty$ are semantically restricted to bind to infinite sets only.

*Remark 1.* It is folklore that languages (without data) accepted by Büchi automata are precisely languages expressible in formulae of the form:

$$\exists_\infty R_1 \cdots \exists_\infty R_m \exists S_1 \cdots \exists S_\ell \ \varphi$$

for some $\varphi \in \mathsf{FO}^2(+1)$. However, we have not found an explicit reference for this result in the literature.

The following theorem is a straightforward generalization of Theorem 5.

**Theorem 6.** *Weak Büchi data automata and $\mathsf{E}_\infty \mathsf{MSO}^2(+1, \sim)$ are equivalent in expressive power.*

*Proof.* The translation from an automaton to a formula uses one additional relation symbol $R$ which is quantified by an $\exists_\infty$ symbol. The formula $\varphi_{accept}$ used in the proof of Lemma 3 is then replaced by

$$\forall x \ (R(x) \to \bigvee_{q \in F} R_q(x)).$$

For the opposite translation, it can be checked with the help of the Büchi condition that relations quantified with $\exists_\infty$ are indeed infinite. □

**Theorem 7.** *The nonemptiness problem for weak Büchi data automata is decidable in 2-NEXPTIME.*

*Proof.* We show in the following that the nonemptiness problem for WBDA can be polynomially reduced to the nonemptiness problem for WDA. The result then follows from Theorem 2. The approach is a classical one. We show that if the language of a WBDA $(\mathcal{A}, \mathcal{C})$ is non-empty then a finite data string of the form $uv$ can be constructed such that there is a run of $\mathcal{A}$ which loops over $v$. The "unravelling" $uv^\omega$ is then also accepted by the automaton. However, some care is needed to assign data values in a suitable manner.

Let $(\mathcal{A}, \mathcal{C})$ be a WBDA with $\mathcal{A} = \langle \Sigma, \Gamma, Q, q_0, \Delta, F \rangle$. Since we are only interested in whether $\mathcal{L}^\omega(\mathcal{A}, \mathcal{C}) = \emptyset$, we can assume, without loss of generality, that the transitions of $\mathcal{A}$ are all of the form $(q, \gamma, q', \gamma)$. Otherwise, we can replace it by a transducer which reads $\Gamma$-strings and guesses, for every position $i$, a symbol $a_i \in \Sigma$, its profile symbol $s_i$ (and store them in the state) and verifies that its output would be (the actual input symbol) $\gamma_i$. Therefore, we consider $\mathcal{A}$ in this proof just as a normal Büchi automaton that gets a $\Gamma$-string as input. The constraints are applied to the same string.

We first fix some notation. We refer to the symbols that occur in key constraints of $\mathcal{C}$ as *key symbols*.

A *zone* is a finite data string over $\Gamma$ in which all positions carry the same data value. An *$\omega$-zone* is an infinite data string over $\Gamma$ in which all positions carry the same data value. The *zones of a data string* $w$ are the maximal zones of $w$. An *adorned zone* is a zone together with a pair $(q, q')$ of states of $\mathcal{A}$. We write $\mathsf{a\text{-}Proj}(z)$ for the triple $(\mathsf{Str}(z), q, q')$ of a zone $z$ that is adorned with the pair $(q, q')$.

We next define an important notion for this proof, (singular and non-singular) witnesses. We will show that the nonemptiness of $(\mathcal{A}, \mathcal{C})$ boils down to deciding whether such witnesses exist. Singular witnesses correspond to data strings in $\mathcal{L}^\omega(\mathcal{A}, \mathcal{C})$ with an infinite zone whereas non-singular witnesses correspond to data strings with finite zones only.

A *singular witness* for $(\mathcal{A}, \mathcal{C})$ is a data string $uv$ over $\Gamma$ the following properties.

- $uv \models \mathcal{C}$.
- There is a state $\hat{q} \in F$ and a (partial) run $\rho = \rho_u \rho_v$ of $\mathcal{A}$ on input $\mathsf{Profile}(uv)_\top$ in which the state after reading $u$ and after reading $v$ is $\hat{q}$. Here, $\mathsf{Profile}(uv)_\top$ denotes the profile string that is obtained from $\mathsf{Profile}(uv)$ by setting the last profile symbol to $\top$.
- All positions of $v$ and the last zone of $u$ carry the same data value and $v$ does not carry any key symbol.

A *non-singular witness* for $(\mathcal{A}, \mathcal{C})$ is a data string $uv$ over $\Gamma$ which fulfills the following conditions.

- All zones in $uv$ are of length at most $|Q|(|\Gamma| + 1)$.
- The data value of the last position of $u$ is different from the value of the first position of $v$.
- There is a state $\hat{q}$ and a (partial) run $\rho = \rho_u \rho_v$ of $\mathcal{A}$ on input $uv$ in which the state after reading $u$ and after reading $v$ is $\hat{q}$. Furthermore, $\rho_v$ contains some state from $F$. In the following, each zone $z$ of $w$ is adorned by the pair $(q, q')$ where $q$ is the state of $\rho$ before reading $z$ and $q'$ is the state after reading $z$.

- The classes of $uv$ can be colored[4] with the four colors black, yellow, white and blue such that all black, yellow and white classes satisfy [5] all constraints from $\mathcal{C}$ and furthermore the following conditions hold.
(black) There are at most $3|Q|^2$ black classes. There are no key symbols in black zones of $v$. Furthermore, it is not the case that the first zone and the last zone of $v$ are from the same black class.
(yellow) There are at most $|Q|^2$ yellow classes and they consist of at most $|\Gamma|$ zones. All these zones are located in $v$.
(white) All zones of the white classes are located in $u$.
(blue) For each blue zone $z$ there is a yellow zone $z'$ such that a-Proj$(z)=$a-Proj$(z')$.

The proof of decidability of the nonemptiness problem for WBDA now reduces to proving the following three claims.

(Claim 1) If there exists a witness for $(\mathcal{A}, \mathcal{C})$ then $\mathcal{L}^\omega(\mathcal{A}, \mathcal{C}) \neq \emptyset$.
(Claim 2) If $\mathcal{L}^\omega(\mathcal{A}, \mathcal{C}) \neq \emptyset$ then there exists a witness for $(\mathcal{A}, \mathcal{C})$.
(Claim 3) There is a nondeterministic algorithm which constructs, for every WBDA $(\mathcal{A}, \mathcal{C})$, in polynomial time some WDA $(\mathcal{A}', \mathcal{C}')$ such that every possible $(\mathcal{A}', \mathcal{C}')$ accepts only witnesses for $(\mathcal{A}, \mathcal{C})$ and for each witness $uv$ there is a run of the algorithm producing some $(\mathcal{A}', \mathcal{C}')$ that accepts $uv$.

Therefore, the nonemptiness problem for WBDA can indeed be reduced non-deterministically in polynomial time to the nonemptiness problem for WDA.

Next, we prove Claims 1-3.

We start by proving Claim 1. Let us assume there is a singular witness $uv$ for $(\mathcal{A}, \mathcal{C})$ (where all names are chosen as above). It is easy to see that in this case $\rho_u \rho_v^\omega$ is an accepting run of $\mathcal{A}$ on input $uv^\omega$ and that $uv^\omega$ satisfies all constraints from $\mathcal{C}$. It should be noted that in $uv^\omega$ all positions in the $v^\omega$-part and some non-empty suffix of $u$ carry the same data value and thus constitute one infinite zone. The repetition of $v$ does not introduce any violations of $\mathcal{C}$ as $v$ does not contain any key symbols.

We next consider the case that $uv$ is a non-singular witness for $(\mathcal{A}, \mathcal{C})$ (where again all names are chosen as above). In principle, we aim again for a data $\omega$-string in $\mathcal{L}^\omega(\mathcal{A}, \mathcal{C})$ that is obtained from $uv$ by repeating $v$ infinitely often. Indeed, by doing so, we obtain a data $\omega$-string whose adorned projection is just a-Proj$(u)$a-Proj$(v)^\omega$. However, the data values cannot be the same in every copy of $v$ as otherwise constraints might be violated.

The basic idea for the assignment of data values is as follows. As white zones only appear in $u$ they are not affected and we do not need to adapt them. As the black zones in $v$ do not contain any key symbols, we can leave them unchanged in each of the infinitely many copies of $v$ that constitute $w$. It only remains to assign data values to the blue zones in $u$ and to the blue and yellow zones in $v$ and the copies of $v$. To this end, we intuitively use the yellow classes as templates. More precisely, we make sure that for every new class that is constituted by assigning data values, the set of zones corresponds to one of the yellow classes of $v$, that is, it has the same set of (adorned) zones as that class. This ensures that each new class satisfies $\mathcal{C}$.

---

[4] Each class gets exactly one color. We refer to zones and positions in a black class as black zones and positions, respectively, and likewise for the other colors.
[5] We do not require that the blue classes satisfy $\mathcal{C}$.

We now describe the construction of the data $\omega$-string $w$ in more detail.

- Let $w_1 = uv^\omega$. Clearly, $\rho_u \rho_v^\omega$ is an accepting run on $\mathrm{Profile}(uv^\omega)$.
- In the remainder of the construction, only data values are changed, but zone projections and runs remain the same.
- Let $w_2$ be a copy of $w_1$ where data values of blue and yellow zones are removed (and thus the black and white zones are just as in $u$ and $v$). As the last zone of $v$ is not from the same black class as the first zone of $v$ it cannot happen that two black zones with the same data value become adjacent by repeating $v$.
- Next, we choose an infinite sequence $d_1, d_2, \ldots$ of data values that do not occur in black or white zones. We assign data values to the blue and yellow zones in $w$ by repeating the following procedure from left to right. In the $i$-th application we constitute a new class by assigning the data value $d_i$ to a set of zones that corresponds to some yellow class.
  - We pick the first yellow or blue zone $z$ that does not yet have a data value. We choose a yellow class $c$ of $v$ that contains a zone $z'$ with $\mathsf{a\text{-}Proj}(z)=\mathsf{a\text{-}Proj}(z')$. This is possible as $z$ is either such a zone itself or it was a blue zone in $uv$ and thus such a zone $z'$ exists by the requirements for blue zones. Let $z'_1, \ldots, z'_k$ be the other zones of the class $c$. For each $i \leq k$ we choose some zone $z_i$ of $w_2$ that has not yet received a data value and fulfills $\mathsf{a\text{-}Proj}(z'_i)=\mathsf{a\text{-}Proj}(z_i)$. We require that the zones $z, z_1, \ldots, z_k$ are pairwise not adjacent. As each yellow zone of $v$ is copied infinitely often in $w$, such zones $z_1, \ldots, z_k$ do exist. We assign to $z, z_1, \ldots, z_k$ the data value $d_i$. We note that the new class has exactly the same zone profile as the yellow class $c$ and therefore satisfies all constraints.

  We denote the resulting data $\omega$-word by $w$.
  It remains to show that indeed $w \in \mathcal{L}^\omega(\mathcal{A}, \mathcal{C})$. As $\rho$ is an accepting run yielding $w$ it only remains to show that all classes of $w$ satisfy $\mathcal{C}$.
  - As white classes have not been changed at all, they clearly satisfy $\mathcal{C}$.
  - As the black zones in $v$ do not contain any key symbols, repeating them in $w$ does not introduce any violations of key constraints. Otherwise, the repetition does not change the set of occurring symbols for any black class and therefore also the inclusion and denial constraints remain valid.
  - Each other class of $w$ has the same set of profiles as some yellow class $c$ of $uv$. Thus, it satisfies all constraints from $\mathcal{C}$ just as $c$ does.

This concludes the proof of Claim 1.

For the proof of Claim 2 let $\mathcal{L}^\omega(\mathcal{A}, \mathcal{C}) \neq \emptyset$ and let $\rho$ be an accepting run on the data $\omega$-word $w$ with $w \models \mathcal{C}$. We consider two cases, depending on whether the number of zones in $w$ is finite or infinite. We first consider the simpler case, in which the number of zones in $w$ is finite. In this case $w = u'v'$, for some finite data string $u'$ and an infinite data string $v'$ such that

- all positions of $v'$ have the same data value $d$,
- there are no key symbols in $v'$.

The latter can be achieved as there is only a finite number of key symbols in the infinite zone.

As $\rho$ is accepting, some accepting state $\hat{q}$ occurs infinitely often in the run $\rho$ on $v'$. Let $u$ be the prefix of $u'v'$ until (and including) the position of $v'$ after which $\hat{q}$ occurs for the first time. Let $v$ be the substring of $v'$ from the first to the second occurrence of $\hat{q}$. Clearly, $uv$ is a singular witness for $(\mathcal{A}, \mathcal{C})$.

Now we turn to the case, where $w$ has an infinite number of zones and therefore all zones are of finite length. The construction of $u$ and $v$ consists of a number of transformation steps of $w$. We refer to the data string obtained after the $i$-th transformation step as $w_i$. In these transformations we intuitively view each $w_i$ as an infinite sequence of zones. We might replace zones by other zones but we never change the sequence of states that $\rho$ takes on the sequence of zone borders. We call the sub-sequence of states that a run $\rho$ takes at zone borders, the *zone sub-run* of $\rho$

Thus, our transformations do not change the zone sub-run of our accepting run. Without loss of generality, we assume that whenever $\mathcal{A}$ assumes a state from $q$ after some position $p$ it also assumes an accepting state after the next symbol with profile $\bot$. This can be accomplished by an easy modification of $\mathcal{A}$. And now we can be sure that in the zone sub-run some infinite state occurs infinitely often.

The first transformation step transforms $w$ into a data string $w_1$ in which each zone has length at most $|Q|(|\Gamma|+1)$. This step is applied to each zone $z$ independently. If $|z| \leq |Q|(|\Gamma|+1)$, nothing has to be done. Otherwise, we mark a set of positions of $z$ such that the first and last position are marked and, for every symbol $\gamma$ that occurs, one occurrence is marked. Thus, at most $|\Gamma|+1$ positions are marked. As $|z| > |Q|(|\Gamma|+1)$ there must be a sequence of at least $|Q|$ unmarked positions in $z$. We consider the state of $\rho$ before the first of these positions and after each of them. Clearly, in this sequence of states some state $q$ must occur twice. By removing the data string between the two occurrences of $q$ we obtain a shorter zone $z'$ with the same set of symbols. Furthermore $z'$ can be obtained by a partial run of $\mathcal{A}$ with the same first and last state as for $z$ in $\rho$, thus the zone sub-run does not change. We note that by removing symbols no key conflicts can be introduced. The repeated application of this process to each zone yields a data word $w_1$ that is accepted by a run of $\mathcal{A}$ with the same zone sub-run as before and for which $w_1 \models \mathcal{C}$.

We select, for each class $c$ of $w_1$, and each symbol $a$ that occurs in $c$, one zone $z$ of $c$ that contains $a$. We call these selected zones the *core zones* of $c$ and the other zones *redundant zones*. Clearly, each class has at most $|\Gamma|$ core zones and remaining zones do not contain any key symbols. Thus, if a redundant zone is removed from a class or a copy of a redundant zone is added to a class the validity of constraints is not affected.

In a nutshell, the remaining transformation steps do the following. First of all, we collect all redundant zones in a finite number of classes. These will be the black classes and they are the only classes that might have an infinite number of zones. From the remaining classes we first distinguish those that contain a zone adornment $(q, q')$ that occurs only a finite number of times. These will be the white classes and there are only finitely many of them. The remaining classes consist only of core zones with adornments that occur infinitely often. We single out a polynomial number of such classes, the yellow classes, that cover all "infinite adornments" and in all remaining classes, the blue ones, we replace all zone strings by strings from yellow zones, thereby ensuring that there exists only a polynomial number of different zone strings outside black and white classes. We now continue the detailed description of the construction.

In the next step, we transform $w_1$ into a data string $w_2$ in which at most $3|Q|^2$ classes have redundant zones. Thus, in particular, at most $3|Q|^2$ classes have infinitely many zones. To this end, we proceed as follows, for every pair $(q, q')$ of states of $\mathcal{A}$. If $(q, q')$ occurs as adornment of any redundant zone of $w_1$ we pick the (up to) first three classes that contain such zones. We color all these classes black.

Next, we modify all redundant[6] zones $z$ that are not (yet) in a black class in a left-to-right fashion as follows. Let $(q, q')$ be the adornment of such a zone $z$. As $z$ is not black there must be three black classes $c_1, c_2, c_3$ with $(q, q')$-adorned redundant zones $z_1, z_2, z_3$, respectively. We replace $z$ by one of $z_1, z_2, z_3$ that has a different data value from the zones adjacent to $z$. Although, this step might change the string projections of zones, the resulting data string still has the same zone sub-run and is therefore still accepted by $\mathcal{A}$. Furthermore, as only redundant zones were removed from non-black classes, these classes still satisfy $\mathcal{C}$. And, as in black classes, only copies of redundant zones are added, they also still satisfy $\mathcal{C}$.

We call a state pair $(q, q')$ *frequent* if it occurs as adornment of infinitely many non-black zones, otherwise *infrequent*. Clearly, there is only a finite number of classes that contain zones with infrequent adornment. We color these classes white.

If all classes are black or white then the construction is finished. Otherwise, the adornment of all zones that are neither black nor white is frequent.

The main goal of the final transformation step is to reduce the number of different string projections that occur in zones that are neither black nor white. We note that this transformation step might cause violations of constraints for blue or yellow classes (but this does not matter as long as we yield a witness). In the following, we choose three positions $p_1, p_2, p_3$ such that one of them ($p \in \{p_1, p_2\}$) will mark the end of $u$ and such that $v$ will (basically) be the data string between $p$ and one of the other two ($p' \in \{p_2, p_3\}$). We pick three such positions to ensure that the condition on the first and last zone of $v$ holds.

Let $\hat{q}$ be some accepting state that occurs infinitely often in the zone sub-run. Such a state exists by our assumption that the automaton assumes accepting states at the end of a zone if it assumed one inside the zone. Let $p_1$ be the minimal position in which $\rho_2$ assumes $\hat{q}$ at the end of a zone and such that all white zones are before $p_1$. We next choose, for each frequent pair $(q, q')$ one class $c_{q,q'}$ of $w_2$ that is neither black nor white, contains a zone with adornment $(q, q')$ and is located after $p_1$. Let $p_2$ be the minimal position in which the zone sub-run assumes $\hat{q}$ and such that all zones of classes $c_{q,q'}$ are before $p_2$.

Now, we choose, for each frequent pair $(q, q')$ one class $c'_{q,q'}$ of $w_2$ that is neither black nor white, contains a zone with adornment $(q, q')$ and is located after $p_2$. Finally, we let $p_3$ be the minimal position in which the zone sub-run assumes $\hat{q}$ and such that all zones of classes $c_{q,q'}$ are before $p_3$.

If the first zone after $p_1$ has a different data value than the last zone before $p_2$ or at least one of them is not black we set $p = p_1$ and $p' = p_2$. Otherwise, if the first zone after $p_2$ has a different data value than the last zone before $p_3$ or at least one of them is not black we set $p = p_2$ and $p' = p_3$. Otherwise, we set $p = p_1$ and $p' = p_3$. In either case, the first zone after $p$ has a different data value than the last zone before $p'$ or at least one of them is not black.

---

[6] We remind the reader that the term redundant is always relative to a class. We note that the zones $z$ are redundant in their original class and also in their black target class.

If $p = p_1$ we color the classes $c_{q,q'}$ yellow, otherwise we color the classes $c'_{q,q'}$ yellow.

In the last transformation step, we color all not yet colored zones in blue and furthermore modify blue zones as follows. Let $z$ be a blue zone with adornment $(q, q')$. As $z$ is neither white nor black, $(q, q')$ is frequent. Let $z'$ be the $(q, q')$-adorned zone in $c_{q,q'}$ (or in $c'_{q,q'}$ if $p = p_2$). We keep the data value of $z$ but replace its string projection by $\mathsf{Str}(z')$. This does not affect the zone sub-run, but it might cause a constraint conflict for the blue class (but, as already mentioned, we need not care about this).

Let $w_3$ be the resulting data string.

Now we define $u$ to be the prefix of $w_3$ until position $p$ and $v$ to be the data string from (excluding) position $p$ to position[7] $p'$.

This construction guarantees that $uv$ is a non-singular witness for $(\mathcal{A}, \mathcal{C})$. This completes the proof of Claim 2.

Finally, we prove Claim 3. As we assume that $\mathcal{A}$ copies its input string to the output (and thus basically is an automaton) the reduction also constructs WDA with this property. We first show how to compute a WDA $(\mathcal{A}', \mathcal{C}')$ for singular witnesses for $(\mathcal{A}, \mathcal{C})$. The algorithm first guesses a symbol $a_0$ that occurs in the infinite class.

$\mathcal{A}'$ has input alphabet $\Gamma \times \{0, 1, 2\}$. The symbols of the form $(a, 1)$ and $(a, 2)$ are used for the class of $v$ and the others for the remaining classes. $\mathcal{A}'$ simulates $\mathcal{A}$ when it reads only the $\Gamma$-part. Furthermore, it guesses a position $p$ (intuitively: the border between $u$ and $v$) that has a state $\hat{q}$ from $F$ and verifies that the final state is $\hat{q}$ as well. For the simulation of the last step $\mathcal{A}'$ behaves as if the final symbol carried a $\top$-symbol in its profile part. $\mathcal{A}'$ further checks that all symbols from $\Gamma \times \{0\}$ occur before position $p - 1$ and that no symbols $(a, 1)$ or $(a, 2)$, where $a$ is a key symbol of $\mathcal{C}$ occur after $p$. It furthermore checks that there is only one occurrence of $(a_0, 2)$ and no other $(a, 2)$. All constraints from $\mathcal{C}$ are reproduced in $\mathcal{C}_1$ separately, for symbols of the form $(a, 0)$ and $(a, 1)$. Furthermore, they ensure that in each class either only $(a, 0)$-symbols occur or only $(a, 1)$-symbols and $(a_0, 2)$ (by denial constraints). Finally, there is an inclusion constraint $V((a, 1)) \subseteq V((a_0, 2))$, for every $a \in \Gamma$ and a key constraint for $(a_0, 2)$, making sure that there is only one class with symbols from $\Gamma \times \{1\}$.

It remains to show how to (non-deterministically) compute a WDA $(\mathcal{A}', \mathcal{C}')$ for non-singular witnesses for $(\mathcal{A}, \mathcal{C})$. The basic idea is that the algorithm first guesses the adorned zones that are used in yellow classes in the order in which they appear in the witness. Furthermore, for each black class, it guesses some symbol $a$ occurring in that class and it guesses the order in which these symbols occur. These symbols are colored with black$'$ instead of black. We use these non-deterministic guesses in the reduction as otherwise, $\mathcal{A}'$ would need to handle, e.g., all possible orders in which the yellow zones appear. This would result in an exponential blow-up.

---

[7] Formally, the position $p'$ might have been modified during the last transformation step. However, we refer by $p'$ to the last position of the zone that corresponds to the zone that ended in $p'$ in $w_2$.

$\mathcal{A}'$ uses the alphabet $\Gamma \times \{\text{black}, \text{black}', \text{yellow}, \text{white}, \text{blue}\} \times \{0, \ldots, 3|Q|^2\}$. It reads colored symbols and always simulates $\mathcal{A}$ on the uncolored projection. It guesses a position $p$ and a state $\hat{q}$ and checks that

- after position $p$ the run has state $\hat{q}$ and likewise at the end;
- between position $p$ and the end, the run assumes some accepting state;
- white symbols only occur before $p$;
- the yellow zones all appear after $p$ and they exactly correspond to the adorned zones that were guessed before;
- each blue zone has the same adorned projection as some yellow zone;
- white, yellow and blue symbols carry a 0 in their last component;
- black symbols carry a non-zero number in their last component;
- there are no black key symbols after $p$;
- it is not the case that the first zone after $p$ is in the same class as the last zone and that they are both black;
- each expected black$'$ symbol occurs exactly once and they occur in the expected order.

In $\mathcal{C}'$ the constraints of $\mathcal{C}$ are reproduced for the black, yellow and white class. Some constraints are added that ensure that in each black class a black$'$-symbol occurs, similarly as for the singular case. Furthermore, in black classes, all symbols have the same number in their last component. In this way, it is ensured that for each $i \in \{1, \ldots 3|Q|^2\}$, there is at most one black class.

Clearly, $\mathcal{A}'$ and $\mathcal{C}'$ can be computed in polynomial time. The computation is deterministic, once the above mentioned values are guessed.

This completes the proof of Claim 3 and thus the proof of the theorem. □

## 6  Conclusion

We conclude this paper with two open problems for future directions. An obvious open problem is the exact complexity of the nonemptiness problem for weak data automata. The current 2-NEXPTIME yields a 3-NEXPTIME upper bound for the satisfiability problem for $\mathsf{EMSO}^2(+1, \sim)$. However, as it is known that this problem can be solved in 2-NEXPTIME [18], some room for improvement is left.

Another interesting question is how our results can be applied to temporal logics. In [10], a restriction of LTL with one register, *simple* LTL, was considered with the same expressive power as some two variable logic. We conjecture that there is a correspondence between our logics and the restriction of simple LTL to the operators $X$, $X^{-1}$ and an operator that allows navigation to some other position.